\documentclass[prb,aps,twocolumn,amsmath,amssymb,floatfix,
superscriptaddress]{revtex4}
\usepackage[dvips]{graphics}
\usepackage{color}
\definecolor{dred}{rgb}{0,0,0.6}
\usepackage{graphicx}  
\usepackage{dcolumn}   
\usepackage{bm}        
\usepackage{amssymb}   
\usepackage{amsmath}
\usepackage{braket}
\usepackage[mathscr]{euscript}
\usepackage{color}
\usepackage{hyperref}
\begin{document}

\title{Phenomenon of multiple reentrant localization in a double-stranded helix with transverse electric field}
\author{Sudin Ganguly}
\email{sudinganguly@gmail.com}
\affiliation{Department of Physics, School of Applied Sciences, University of Science and Technology Meghalaya, Ri-Bhoi-793101, India}

\author{Suparna Sarkar}
\email{physics.suparna@gmail.com }
\affiliation{Physics and Applied Mathematics Unit, Indian Statistical
  Institute, 203 Barrackpore Trunk Road, Kolkata-700108, India}

\author{Kallol Mondal}
\email{kallolsankarmondal@gmail.com}
\affiliation{School of Physical Sciences, National Institute of Science Education and Research Bhubaneswar, Jatni, Odisha 752050, India}
\affiliation{Homi Bhabha National Institute, Training School Complex, Anushaktinagar, Mumbai 400094, India}

\author{Santanu K. Maiti}
\email{santanu.maiti@isical.ac.in}
\affiliation{Physics and Applied Mathematics Unit, Indian Statistical
  Institute, 203 Barrackpore Trunk Road, Kolkata-700108, India}


\begin{abstract}
The present work explores the potential for observing multiple reentrant localization behavior in a double-stranded helical (DSH) system, extending beyond the conventional nearest-neighbor hopping interaction. The DSH system is considered to have hopping dimerization in each strand, while also being subjected to a transverse electric field. The inclusion of an electric field serves the dual purpose of inducing quasiperiodic disorder and  strand-wise staggered site energies. Two reentrant localization regions are identified: one exhibiting true extended behavior in the thermodynamic limit, while the second region shows quasi-extended characteristics with partial spreading within the helix. The DSH system exhibits three distinct single-particle mobility edges linked to localization transitions present in the system. The analysis in this study involves examining various parameters such as the single-particle energy spectrum, inverse participation ratio, local probability amplitude, and more. Our proposal, combining achievable hopping dimerization and induced correlated disorder, presents a unique opportunity to study phenomenon of reentrant localization, generating significant research interest.
\end{abstract}


\maketitle

\section{\label{sec1}Introduction}
The phenomenon of localization has been a vibrant area of research in condensed matter physics ever since its prediction by P. W. Anderson~\cite{al1}. Over the years, the interest in this topic has grown exponentially with the exploration of various fascinating systems across different branches of physics~\cite{al2,al3,al4,al5,al6,al7,al8}. Anderson's seminal work~\cite{al1} demonstrated a metal-insulator transition in a one-dimensional atomic system with uncorrelated site energies, where all energy eigenstates become completely localized regardless of the strength of disorder. However, such an uncorrelated disordered system is considered relatively trivial due to the absence of a finite critical disorder strength and the limited control over it. By imposing constraints on the site energies, one can unveil captivating dynamics and explore more intriguing phenomena within correlated disordered systems~\cite{cd1,cd2,cd3,cd4,cd5}.

To date, a wide range of correlated systems have been employed across various fields, and among them, the Aubry-Andr\'{e}-Harper (AAH) model~\cite{aah1,aah2} stands out as the most prevalent and adaptable example. In the nearest-neighbor tight-binding (TB) framework, the 1D AAH model with an incommensurate potential demonstrates a distinct transition between localization and delocalization. Below a critical point, all eigenstates are found to be delocalized, while beyond that critical point, they become completely localized~\cite{aah1,aah2,gaah1}. Recent advancements in the field have introduced several generalizations of this model. These include exponential short-range hopping~\cite{gaah1}, flatband networks~\cite{gaah2}, higher dimensions~\cite{gaah3}, power-law hopping~\cite{gaah4}, flux-dependent hopping~\cite{gaah5}, and nonequilibrium generalized AAH model~\cite{gaah6}, etc. The studies have revealed that beyond the nearest-neighbor TB framework, there is typically a single-particle mobility edge (SPME), which represents a critical energy that differentiates localized states from extended states within the system~\cite{mott}. AAH systems have also been experimentally realized using cold atoms and optical waveguides.~\cite{exaah1,exaah2}.

Based on current understanding, it has been firmly established that following a localization transition, all states continue to exhibit localization indefinitely as the disorder strength increases. However, recent studies have revealed that under certain constraints or conditions imposed on the system, this characteristic of indefinite localization may alter. In more recent findings, an intriguing occurrence of reentrant localization has been discovered in 1D quasiperiodic disordered systems~\cite{reentrant1,reentrant2,reentrant3,reentrant4,reentrant5,reentrant6,reentrant7,reentrant8,reentrant9,reentrant10}, which can be attributed to the interplay between hopping dimerization and the presence of staggered AAH disorder. In this localization phenomenon, as the strength of the staggered potential increases, certain states that were previously localized undergo a transition and regain their extended character. Up to this point, reentrant localization has paved the way for obtaining a comprehensive understanding of Anderson localization theory. As a consequence, it has generated significant interest from both the theoretical and experimental domains. {\it Building upon this line of research, we introduce a methodology to observe the reentrant behavior in systems that go beyond the nearest-neighbor hopping scenario}. 

Specifically, we focus on a DSH system and investigate its response when exposed to a transverse electric field. Interestingly, the application of a transverse electric field has the intriguing capability to introduce AAH disorder as well as the staggered scenario into the helical system~\cite{helec1,helec2}. This makes the system highly desirable from an experimental standpoint. In the context of the geometrical structure of a helical system, the nature of hopping interactions can vary, being either short-range or long-range. However, recent research~\cite{reentrant6} suggests that long-range hopping can weaken the competition between dimerized hopping and the staggered potential, leading to the absence of reentrant behavior. Considering this insight, we focus solely on short-range hopping interactions in our current work. By exclusively examining short-range hopping, we are able to observe and validate the presence of multiple localization phenomena using various analytical techniques. These techniques include analyzing the eigenvalue spectrum, inverse participation ratio, and local probability amplitudes, among others. Through these investigations, we gain valuable insights into the behavior and characteristics exhibited by the system.  

The key findings of the present work are: (i) presence of multiple reentrant behavior, specifically two instances of reentrant localizations, (ii) states in the first reentrant region exhibit truly extended nature in the thermodynamic, (iii) states in the second reentrant region are quasi-extended, and (iv) implementation of reentrant phenomenon in a realistic biological system simply by applying an electric field.   

The rest of the paper is organized as follows. In Sec.~\ref{sec2}, we describe the helical geometry, the tight-binding Hamiltonian in presence of transverse electric field, and the theoretical formulae for the quantities required to study the localization phenomenon. The numerical results and our analysis are presented in Sec.~\ref{sec3}. Finally, in Sec.~\ref{conclusion}, we conclude our findings. 

\begin{figure}[h] 
\centering
\includegraphics[width=0.35\textwidth]{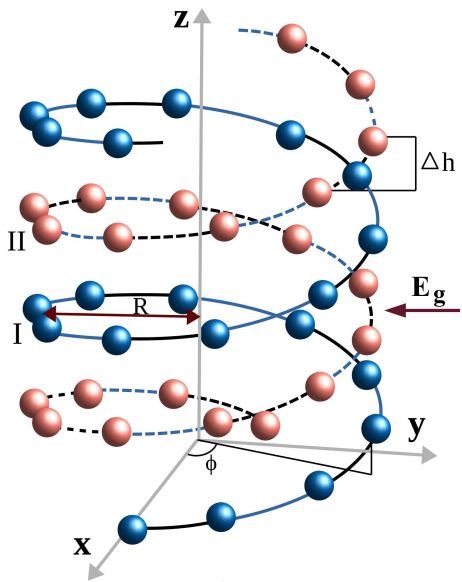}
\caption{(Color online). Schematics of a right-handed double-stranded helical geometry in presence of an external electric field of strength $E_g$. The blue balls represent the sites in strand-I, and the red balls represent the sites in strand-II. $R$ is the radius of the helix and $\Delta h$ is the stacking distance between adjacent sites. $\phi = n\Delta \phi$, where $\Delta \phi$ is the twisting angle between the neighboring sites and $n$ is the site index in each strand. The alternating black (dotted black) and blue (dotted blue) lines indicate dimerization of the adjacent hoppings in strand-I (strand-II).}
\label{setup}
\end{figure}

\section{\label{sec2}System and theoretical framework}
Figure~\ref{setup} depicts the schematic diagram of a right-handed double-stranded helical geometry. The alternate bondings are assumed to be dimerized as shown by 
the black (dotted black) and blue (dotted blue) lines in strand-I (strand-II). An electric field $E_g$ is applied perpendicular to the axis of the helix. The nature of hopping is determined by two parameters: the stacking distance $\Delta h$ and the twisting angle $\Delta \phi$. When $\Delta h$ is sufficiently small, indicating densely packed atoms, long-range hopping becomes significant as electrons can effectively hop over larger distances. In contrast, when $\Delta h$ is considerably large, with atoms separated by greater distances, electron motion is restricted to shorter distances, resulting in a short-range hopping helix. In practice, two prominent examples that fall into the short-range hopping and long-range hopping groups are DNA and protein molecules, respectively~\cite{qfsung}. However, in this study, our focus is solely on SRH systems, as mentioned previously.

The tight-binding Hamiltonian for the DSH system in the presence of an external electric field is expressed as

\begin{eqnarray}
H &=&\sum_{j=I,II} \left[\sum_{\substack{n=1\\ \hfill}}^N \epsilon_{j,n} c_{j,n}^\dagger c_{j,n} \right.\nonumber\\
  &+& \left. t_1\sum_{n=1,3,5,...}^N\left(c_{j,n}^\dagger c_{j,n+1} +h.c.\right) \right.\nonumber\\
  &+& \left. t_2\sum_{n=2,4,6,...}^N\left(c_{j,n}^\dagger c_{j,n+1} +h.c.\right)\right.\nonumber\\
  &+& \left.\sum_{n=1}^{N} \sum_{\substack{m=1\\ m-n\neq \pm 1}}^{N} t_{j,(n,m)}\left(c_{j,n}^\dagger c_{j,m} + h.c. \right)\right]\nonumber\\
&+&\sum_n t_3\left(c_{I,n}^\dagger c_{II,n} + h.c.\right).
\label{ham}
\end{eqnarray}

Here $j(=I,II)$ represents the strand index, $c_{j,n}^\dagger$ and $c_{j,n}$ are the usual fermionic creation and annihilation operators at the $n$th site of strand-$j$, respectively. 

In the first term of Eq.~\ref{ham}, $\epsilon_{j,n}$ represents the site energy at site $n$ of strand-$j$. When an electric field is applied perpendicular to the helix axis, the site energy undergoes modifications as~\cite{helec1, helec3, helec4}
\begin{equation}
\epsilon_{I,n} = -\epsilon_{II,n} = eV_g \cos{(n\Delta \phi - \beta)},
\label{ons}
\end{equation}
where $e$ is the electronic charge and $V_g$ corresponds to the gate voltage associated with the applied electric field. The relationship between the gate voltage and the electric field can be expressed as $2 V_g = 2 E_g R$. The reversal in sign observed between the two strands can be attributed to the combined effects of the perpendicular electric field and the helix conformation of the strands~\cite{helec1}. The phase factor $\beta$ represents the angle between the positive $x$-axis and the incident electric field. This phase factor can be adjusted or modified by changing the direction of the electric field. Equation~\ref{ons} illustrates that the presence of a perpendicular electric field leads to a harmonic modulation of the site energies along the helical strands. Interestingly, such a modulation is identical to the well-known AAH model~\cite{aah1,aah2}. The factor $eV_g$ is analogous to the AAH modulation strength $W$, $\Delta \phi$ can be identified with the term $2\pi b$ ($b$ an irrational number) and the phase $\beta$ with the Aubry phase $\phi_\nu$  in the AAH model. By selectively choosing the term $\Delta \phi$, it becomes possible to achieve a deterministic disordered double-stranded helical system, where the site energies exhibit a correlated pattern resembling the AAH model. This correlation is realized when the DSH system is subjected to the electric field $E_g$. 

The second and third terms in Eq.~\ref{ham} represent the nearest-neighbor hopping terms in the Hamiltonian. The parameters $t_1$ and $t_2$ indicate that the hopping in the DSH system is dimerized.

The fourth term in Eq.~\ref{ham} is the beyond nearest-neighbor interaction. $t_{j,(n,m)}$ is the hopping integral between the sites $n$ and $m$ in strand-$j$ and reads as~\cite{helec1,helec5}
\begin{equation}
t_{j,(n,m)} = \left(\frac{t_1 + t_2}{2}\right) {\rm{e}}^{-\left(l_{j,(n,m)} - l_1\right)/l_c},
\label{hop}
\end{equation}
where $l_{j,(n,m)}$ is the Euclidean distance between sites $n$ and $m$. With $n-m=k$, it is expressed as
\begin{equation}
l_{j,(n,m)} = \sqrt{\left[2R\sin{\left(\frac{k\Delta\phi}{2}\right)}\right]^2 + \left[k\Delta h\right]^2},
\label{dist}
\end{equation}
and $l_1$ represents the distance between neighboring sites in both strands, and its value can be calculated using Eq.\ref{dist} when $k=1$. On the other hand, $l_c$ denotes the decay exponent. In Eq.\ref{hop}, the first term within the parentheses, $\left(t_1 + t_2\right)/2$, accounts for an average over a unit cell. This average is utilized in the computation of hopping integrals beyond the nearest-neighbor interactions.

The final term in Eq.~\ref{ham} corresponds to the inter-strand coupling, which describes the interaction between the two strands and $t_3$ represents the inter-strand hopping integral.

The inverse participation ratio (IPR) serves as a valuable tool for detecting the transition from a localized state to a delocalized state. This measure allows us to quantify and analyze the spatial distribution of a wavefunction or probability density, providing insights into whether the state is confined to a specific region or spread out across multiple locations. By observing changes in the IPR, we can effectively identify and track the transition as the wavefunction evolves from a localized state to a more delocalized one or vice-versa. For the $n$th normalized eigenstate, IPR is defined as~\cite{def-ipr1,def-ipr2}
\begin{equation}
\text{IPR}_n = \sum_i \lvert \psi_n^i \rvert^4.
\label{iprn}
\end{equation}
In the case of a highly extended state, in the thermodynamic limit, the IPR tends to zero. On the other hand, for a strongly localized state, the IPR approximately approaches to unity.

A complementary tool to characterize the localization transition is the normalized participation ratio (NPR), which for the $n$th normalized eigenstate is defined as~\cite{def-ipr1,def-ipr2}
\begin{equation}
\text{NPR}_n = \left(2N\sum_i \lvert \psi_n^i \rvert^4\right)^{-1}
\end{equation}
where, $2N$ is the total number of sites present in the DSH system. In the case of a highly extended state, the NPR tends to unity in the thermodynamic limit. Conversely, for a strongly localized state, the NPR approaches to zero.

The earlier defined IPR$_n$ and NPR$_n$ can be modified to characterize the parameter space region where localized and delocalized states coexist. One defines their average over a
subset of states $N_L$ as follows~\cite{def-ipr2}
\begin{equation}
\langle \text{IPR} \rangle = \sum_n^{N_L} \frac{\text{IPR}_n}{N_L}, \hfill \langle \text{NPR} \rangle = \sum_n^{N_L} \frac{\text{NPR}_n}{N_L}.
\label{avg}
\end{equation}
When all $N_L$ states are localized, $\langle \text{IPR} \rangle$ tends to unity, while $\langle \text{NPR} \rangle$ tends to zero when all $N_L$ states are delocalized. However, in the regime where both $\langle \text{IPR} \rangle$ and $\langle \text{NPR} \rangle$ remain finite, the Hamiltonian's spectrum features an intermediate phase with coexisting spatially extended and localized eigenstates, along with the presence of SPME.

\section{\label{sec3}Results and Discussion}
Let us mention the common parameter values before presenting the numerical results. To implement the short-range hopping in DSH system, we consider physical parameters analogous to those found in the real biological system~\cite{dna}. DNA has been proposed as an ideal and established example of short-range hopping system by various research groups. The structural parameters for the said geometry are as follows: The radius is considered as $R=8\,$\AA, the stacking distance as $\Delta h=4.3\,$\AA , the twisting angle $\Delta\phi=\pi\left(\sqrt{5}-1\right)/4$, and the decay exponent $l_c=0.8\,$\AA. From the relation $\Delta\phi = 2\pi b$, we can determine the value of $b$ for the SRH case, which is incommensurate. All the energies are measured in units of eV. The number of sites in each of the strands is taken as $N=500$. Unless stated otherwise, we set the dimerized hopping integrals as $t_1=0.5$, $t_2=2.2$, and the inter-strand coupling $t_3=1$.

\begin{figure}[h] 
\centering
\hskip 0.15 in
\includegraphics[width=0.2\textwidth]{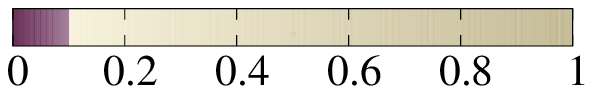}\\
\includegraphics[width=0.24\textwidth]{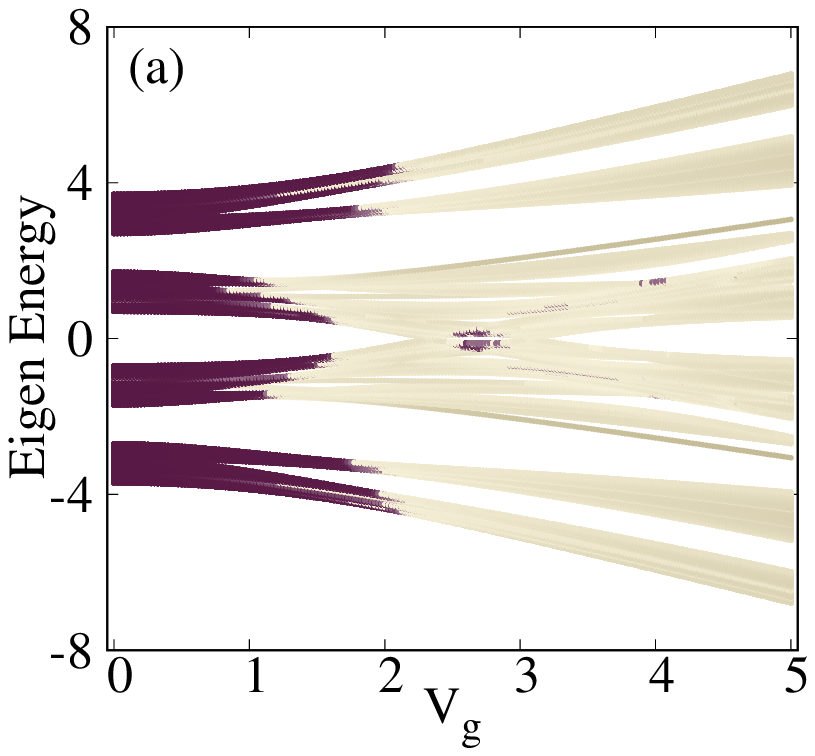}\vskip 0.1 in\includegraphics[width=0.24\textwidth]{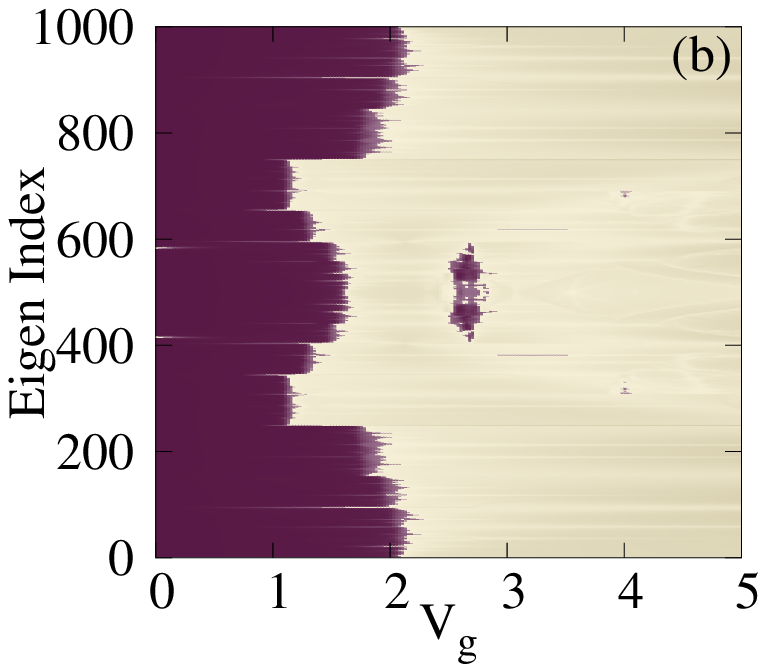}\hfill\includegraphics[width=0.24\textwidth]{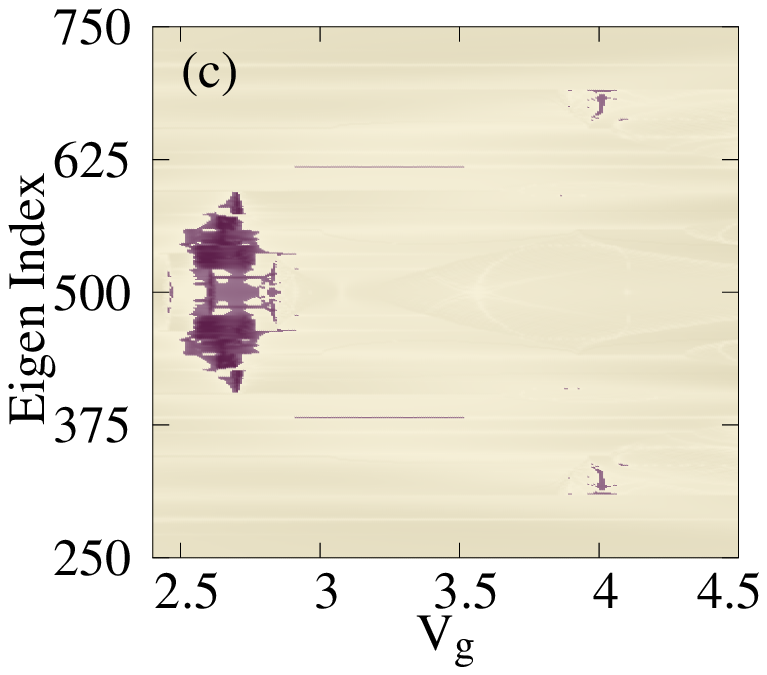}
\caption{(Color online). Density plot. (a) The energy spectrum vs gate voltage $V_g$ with $t_1 = 0.5$, $t_2 = 2.2$, and $t_3=1$. (b) The energy index vs gate voltage $V_g$. (c) A magnified version of Fig.~\ref{wvar}(b) to provide enhanced clarity. The color map shows IPR values of different energy eigenstates.}
\label{wvar}
\end{figure}

To study the localization transition, the energy spectrum corresponding to the Hamiltonian in Eq.~\ref{ham} is plotted as a function of gate voltage $V_g$ (in units of Volts) as shown in Fig.~\ref{wvar}(a). Each energy point in the plot is color-coded based on its corresponding IPR value, which is computed according to Eq.~\ref{iprn}. To capture the localization transition, our colorbar uses purple for the lowest 10$\%$ of the maximum IPR value, highlighting the extended states and a gradient of increasing gray shades for the rest, reflecting higher degree of localization. In Fig.~\ref{wvar}(a), the purple color extends throughout the entire region below $V_g\sim 1$, indicating that the IPR values in this region are significantly below 0.1. This observation strongly suggests that the states within this region exhibit extended behavior. Above $V_g\sim 1$, a mixed phase emerges where some states begin to localize while others remain extended, resulting in a combination of both localized and extended states. This mixed phase persists until approximately $V_g\sim 2$. However, beyond this critical value, all the states undergo localization, indicating a complete transition to a fully localized state. Around $V_g\sim 2.6$, an intriguing phenomenon occurs as a small number of states around zero energy undergo a reentrant localization, indicated by a narrow purple patch within the predominantly localized region. This region exhibits a transient delocalization, where a few states regain their extended nature in contrast to the surrounding localized states. Upon crossing the reentrant zone, 
all the states return to a localized state. However, as we approach $V_g\sim 4$, a noteworthy phenomenon occurs. Several small purple spots emerge, indicating the presence of a second reentrant region. Within this region, a few states exhibit a transient delocalization before ultimately undergoing localization once again. A better visibility of the situation can be obtained by examining the individual eigenstates' IPR, as depicted in Fig.~\ref{wvar}(b). This plot provides a comprehensive view of the localization behavior and allows for a more detailed analysis of the reentrant regions and the transition between extended and localized states. The presence of the first reentrant region, spanning from approximately $V_g\sim 2.5$ to 2.9, is clearly evident in the plot. Within this range, a significant number of eigenstates exhibit delocalization, marked by a distinct decrease in their IPRs. Similarly, the occurrence of the second reentrant region around $V_g\sim 3.9$ to 4.1 is also observed, with a noticeable deviation from the localized behavior as indicated by a cluster of eigenstates displaying lower IPR values. For a more enhanced visualization, a magnified section of Fig.\ref{wvar}(b) is illustrated in Fig.\ref{wvar}(c), providing a clear depiction of the aforementioned description. In Fig.~\ref{wvar}(c), we observe the presence of two horizontal lines highlighted in purple color immediately following the first reentrant localization. To assess the potential occurrence of another reentrant localization, we thoroughly analyze the IPR values and the $V_g$-window associated with these two lines. Upon investigation, it becomes evident that the IPR values associated with these horizontal lines are approximately 0.09, indicating the presence of quasi-extended states. However, it should be noted that these horizontal lines appear before the completion of the first reentrant localization. Consequently, we can conclude that these two lines do not represent another instance of reentrant behavior.

To gain insight into the mixed phase zone, we compute the average IPR and NPR over a subset of states $N_L$ from the spectrum of Fig.~\ref{wvar}, 
\begin{figure}[h] 
\centering
\includegraphics[width=0.35\textwidth]{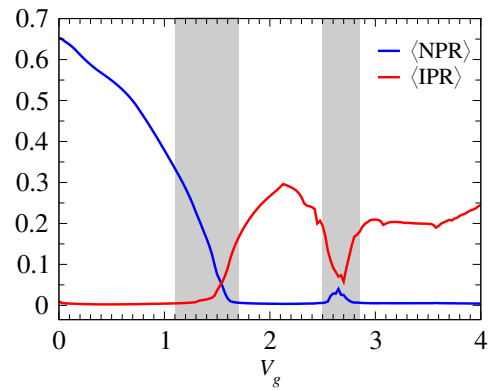}
\caption{(Color online). $\langle \text{IPR} \rangle$ and $\langle \text{NPR} \rangle$ as a function of $V_g$ for a subset of states ranging from 400 to 600 of Fig.~\ref{wvar}(b). The shaded regions indicate the critical zones. All the system parameters remain the same as described in Fig.~\ref{wvar}.}
\label{npr-ipr}
\end{figure}as defined in Eq.~\ref{avg}. The quantities $\langle \text{IPR} \rangle$ and $\langle \text{NPR} \rangle$ are plotted as a function of $V_g$ in Fig.~\ref{npr-ipr}. In this analysis, $N_L$ is considered to be the subset of eigenstates with indices ranging from 400 to 600, taken from Fig.~\ref{wvar}(b). All the system parameters remain unchanged as described earlier. 
In Fig.~\ref{npr-ipr}, both $\langle \text{IPR} \rangle$ and $\langle \text{NPR} \rangle$ exhibit finite values within the range of $1.1 < V_g < 1.7$, indicating the presence of a critical region where a mixture of extended and localized states coexist. For $V_g > 1.7$, the system undergoes a transition into a fully localized state, where all states become localized. Moreover, in the approximate window of $2.5 < V_g < 2.9$, a dip in the $\langle \text{IPR} \rangle$ value is observed, accompanied by a bump in $\langle \text{NPR} \rangle$. This specific region corresponds to the occurrence of the first reentrant region. Within the chosen subset of states, the system hosts two SPMEs. Considering the limited number of extended states in the second reentrant region, detection of the transition becomes challenging within the same plot. Nevertheless, it is important to note that when considering the entire spectrum, the system reveals the presence of three distinct SPMEs.

To explore the extension of states within the reentrant regions, we analyze the local probability amplitudes of different states at varying gate voltages.
\begin{figure}[h] 
\centering
\vskip 0.1 in
\includegraphics[width=0.495\textwidth]{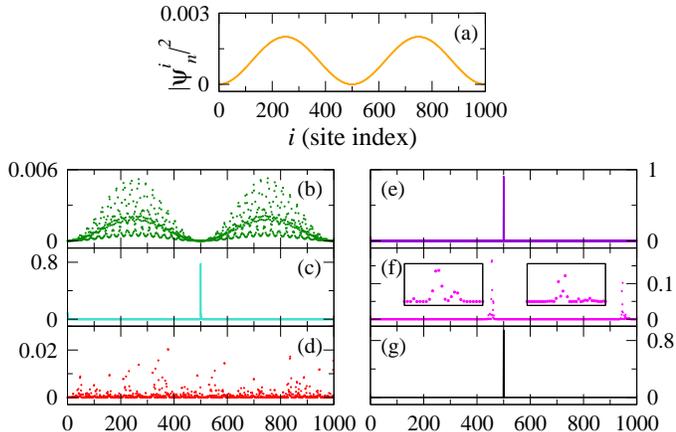}
\caption{(Color online). Local probability amplitude $\lvert \psi_n^i\rvert^2$ vs site index $i$ for the $n$th eigen state. (a) $V_g=0$ and $n=500$, (b) $V_g=1$ and $n=500$, (c) $V_g=2.3$ and $n=249$, (d) $V_g=2.65$ and $n=436$, (e) $V_g=3.5$ and $n=249$, (f) $V_g=4$ and $n=332$, and (g) $V_g=5$ and $n=249$.}
\label{loc-prob}
\end{figure}
This analysis provides insights into the robustness of state extension or localization within the system as the gate voltage, $V_g$ changes. The results are presented in Fig.~\ref{loc-prob}. Firstly, we calculate the local probability amplitude $\lvert \psi_n^i\rvert^2$ for the state $n=500$ under zero-field condition, as illustrated in Fig.~\ref{loc-prob}(a). In this disorder-free case, as expected, the local probability amplitudes $\lvert \psi_n^i \rvert^2$ for all sites exhibit extended behavior. This is evident from the smooth sinusoidal curve and the relatively lower values of probability amplitudes throughout the system. Next, we examine the case where $V_g=1$ and focus on the state $n=500$, with the corresponding result depicted in Fig.~\ref{loc-prob}(b). Notably, the envelope of the local probability distribution maintains the characteristics observed in the disorder-free scenario. Consequently, the state remains within the extended region. Subsequently, we raise the gate voltage to $V_g=2.3$ and examine the state $n=249$. As depicted in Fig.~\ref{loc-prob}(c), the probability amplitudes for all sites, except for site index $i=500$, become vanishingly small. Notably, at this specific site, the probability amplitude assumes a relatively large value of approximately 0.8. This observation indicates that the chosen $V_g$ value indeed induces a fully localized state within the system. In Fig.~\ref{loc-prob}(d), we examine the case where $V_g$ is fixed at 2.65, corresponding to the first reentrant localization region. We consider the state $n=436$, and observe that the probability amplitudes range from 0 to 0.02, indicating relatively low values. Therefore, it is evident that within the first reentrant region, the considered state regains its extended nature. Upon further increasing $V_g$ to 3.5 and examining the state $n=249$, it is evident from Fig.~\ref{loc-prob}(e) that the system transitions into a fully localized phase once again. To investigate the second reentrant localization, we examine the case where $V_g=4$ and focus on the state $n=332$. Interestingly, we observe two broad peaks in the distribution of $\lvert \psi_n^i \rvert^2$, as shown in Fig.~\ref{loc-prob}(f). The values of the probability amplitude for these peaks are relatively low. Upon closer inspection, we find that these peaks are spread over a span of approximately 40-50 sites, as demonstrated in the two insets of Fig.~\ref{loc-prob}(f). Consequently, this region exhibits a quasi-extended behavior. In Fig.~\ref{loc-prob}(g), we set the gate voltage to $V_g=5$ and examine the state $n=249$. Notably, the probability amplitude is localized predominantly at site $n=500$ with a value of approximately 0.8, while the amplitudes at all other sites are vanishingly small. This observation confirms the presence of a fully localized state within the system.

\begin{figure}[h]
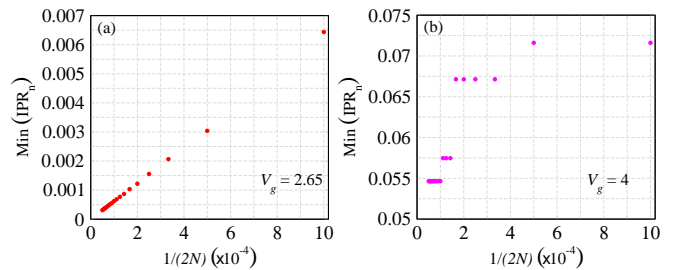
 
\centering
\includegraphics[width=0.24\textwidth]{fig5a.eps}\hfill\includegraphics[width=0.24\textwidth]{fig5b.eps}
\caption{(Color online). Minimimum IPR$_n$ as a function of $1/2N$ in the two reentrant regions, namely, at (a) $V_g=2.65$ and (b) $V_g=4$. }
\label{finite}
\end{figure}

To address and account for any potential finite size effects, we examine the relationship between the minimum IPR value and the system size in the two reentrant regions as shown in Fig.~\ref{finite}. The minimum IPR value for a given system size is determined by identifying the lowest IPR among all states, 
achieved at a specific value of $V_g$. We plot IPR$_n$ as a function of the inverse of the sytem size $1/2N$ in the first reentrant region, namely at $V_g=2.65$ as shown in Fig.~\ref{finite}(a). As the system size increases, the IPR$_n$ value decreases following a scaling behavior of $\mathcal{O}(1/L)$, where $L$ represents the system size. Consequently, in the thermodynamic limit, the states within the first reentrant region display a tendency towards a true extended nature. In contrast, the results shown in Fig.~\ref{finite}(b) for the second reentrant region do not exhibit a scaling behavior similar to the first reentrant region. Instead, the IPR$_n$ value decreases with increasing system size in a step-like fashion. In the limit of large system sizes, it converges to a finite value of approximately 0.055. Considering the lower values of IPR and its behavior with respect to system size, it becomes evident that the states within the second reentrant region do not exhibit a genuine extended nature in the thermodynamic limit. Instead, these states can be characterized as quasi-extended, as observed in Fig.~\ref{loc-prob}(f), where they demonstrate a partial spreading throughout the system.

\begin{figure}[h] 
\centering
\includegraphics[width=0.24\textwidth]{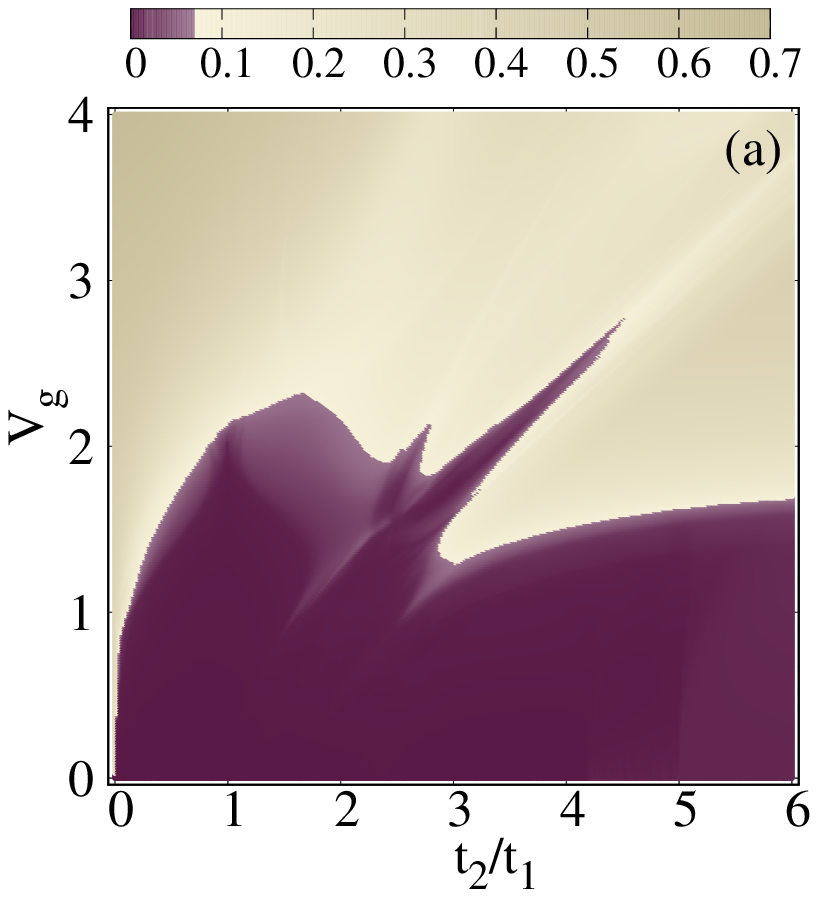}\includegraphics[width=0.24\textwidth]{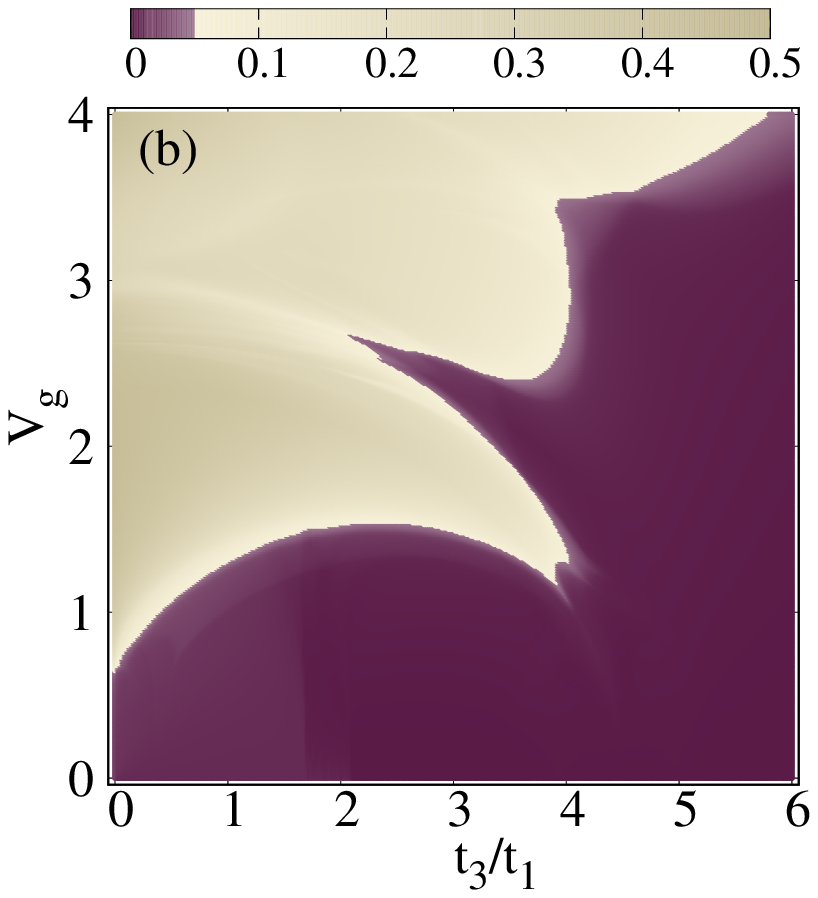}
\caption{(Color online). Density plot. The phase diagram in (a) $t_2/t_1$ and $V_g$ space, (b) $t_3/t_1$ and $V_g$ spce. The color map shows $\langle \text{IPR} \rangle$ value associated with the subset of states mentioned in Fig.~\ref{npr-ipr}. The physical parameters remain the same as in Fig.~\ref{wvar}. }
\label{phase}
\end{figure}

Finally, we examine the parameter space between the gate voltage $V_g$ and the hopping integrals in terms of average IPR $(\langle \text{IPR} \rangle)$ to identify the regions where the phenomenon of reentrant localization emerges. Here, our focus is solely on the first reentrant region, and we do not investigate the second reentrant region due to the aforementioned reasons. In Fig.\ref{phase}(a), we plot the color-coded $\langle \text{IPR} \rangle$ as functions of $V_g$ and $t_2/t_1$. All other physical parameters are the same as described in Fig.\ref{wvar}. We calculate $\langle \text{IPR} \rangle$ using the 
same method as described in Fig.~\ref{npr-ipr}. The maximum IPR value in the corresponding color bar is 0.7, and the extended nature is attributed to the range from 0 to 0.07, which corresponds to $10\%$ of the maximum IPR value. Based on the plot, we observe that the reentrant region emerges for values of $t_2/t_1$ ranging approximately from 2.9 to 4.5 and for $V_g$ within the range of 1.3 to 2.8. By adjusting the inter-strand coupling $t_3$, it is also feasible to alter the extent of the reentrant region. To visualize this, we plot the color-coded $\langle \text{IPR} \rangle$ as functions of $V_g$ and $t_3/t_1$ in Figure~\ref{phase}(b). We observe an approximate reentrant window occurring for $t_3/t_1$ values ranging from 2.2 to 4 and for $V_g$ within the range of 1 to 2.6. It is important to note that the parameter space considered is based on the selected subset of eigenstates, as mentioned earlier. The specific values might slightly vary if we were to consider the entire spectrum.

\section{\label{conclusion}Conclusion}
We have focused on the localization behavior of a DSH system under the influence of a transverse electric field. Each strand of the DSH system is assumed to possess dimerized hopping. The introduction of a transverse electric field gives rise to the emergence of correlated disorder within the system, accompanied by a strand-wise staggered arrangement of site energies. Notably, we have observed a distinctive multiple reentrant behavior, specifically characterized by two instances of reentrant localizations. This observation has been made by examining the behavior of the IPR of individual eigenstates within the single-particle spectrum. Each localization transition is accompanied by an SPME, and our system exhibits a total of three SPMEs, two of which are associated with the two reentrant regions. By examining the local probability amplitude and scaling behavior of IPR, We have found that the states corresponding to the first reentrant region demonstrate genuine extended characteristics in the thermodynamic limit. However, states within the second reentrant region display a quasi-extended nature. Our investigation reveals that the reentrant region can be influenced and adjusted by modulating both the gate voltage and hopping integrals.

Considering the ongoing progress in experimental feasibility to achieve hopping dimerization~\cite{dim1,dim2,dim3,dim4,dim5,dim6} and the potential for inducing correlated disorder (AAH) through a transverse electric field, our proposal is highly compelling and is expected to generate significant interest within the research community. The incorporation of these factors in our study presents a valuable opportunity to observe and study the reentrant localization behavior.

\setcounter{secnumdepth}{0}


\begin{thebibliography}{99}
\bibitem{al1} P. W. Anderson, {\it Absence of diffusion in certain random lattices}, Phys. Rev. \textbf{109}, 1492 (1958).

\bibitem{al2} P. A. Lee and T. V. Ramakrishnan, {\it Disordered electronic systems}, Rev. Mod. Phys. \textbf{57}, 287 (1985).

\bibitem{al3} B. Kramer and A. MacKinnon, {\it Localization: theory and experiment}, Rep. Prog. Phys. \textbf{56}, 1469 (1993).

\bibitem{al4} D. S. Wiersma, P. Bartolini, A. Lagendijk, and R. Righini, {\it Localization of light in a disordered medium}, Nature \textbf{390}, 671 (1997).

\bibitem{al5} T. Schwartz, G. Bartal, S. Fishman, and M. Segev, {\it Transport and Anderson localization in disordered two-dimensional photonic lattices}, Nature \textbf{446}, 52 (2007).

\bibitem{al6} H. Hu, A. Strybulevych, J. H. Page, S. E. Skipetrov, and B. A. van Tiggelen, {\it  Localization of ultrasound in a three-dimensional elastic network}, Nat. Phys. {\bf 4}, 945 (2008).

\bibitem{al7} W. R. McGehee, S. S. Kondov, W. Xu, J. J. Zirbel, and B. DeMarco, {\it Three-Dimensional Anderson Localization in Variable Scale Disorder}, Phys. Rev. Lett. {\bf 111}, 145303 (2013). 

\bibitem{al8} M. Segev, Y. Silberberg, and D. N. Christodoulides, {\it Anderson localization of light}, Nature Photon {\bf 7}, 197 (2013).

\bibitem{cd1} F. A. B. F. de Moura and M. L. Lyra, {\it Delocalization in the 1D Anderson model with long-range correlated disorder}, Phys. Rev. Lett. {\bf 81}, 3735 (1998).

\bibitem{cd2} F. M. Izrailev and A. A. Krokhin, {\it Localization and the Mobility Edge in One-Dimensional Potentials with Correlated Disorder}, Phys. Rev. Lett. {\bf 82}, 4062 (1999).

\bibitem{cd3} P. Carpena, P. Bernaola-Galv\'{a}n, P. C. Ivanov, and H. E. Stanley, {\it Metal-insulator transition in chains with correlated disorder}, Nature {\bf 418}, 955 (2002).

\bibitem{cd4} F. M. Izrailev, A. A. Krokhin, and N. M. Makarov, {\it Anomalous localization in low-dimensional systems with correlated disorder}, Phys. Rep. {\bf 512}, 125 (2012).

\bibitem{cd5} G. M. Conley, M. Burresi, F. Pratesi, K. Vynck, and D. S. Wiersma, {\it Localization and the mobility edge in one-dimensional potentials with correlated disorder}, Phys. Rev. Lett. {\bf 112}, 143901 (2014).

\bibitem{aah1} P. G. Harper, {\it The general motion of conduction electrons in a uniform magnetic field, with application to the diamagnetism of metals}, Proc. R. Soc. London, Ser. A {\bf 68}, 874 (1955).

\bibitem{aah2} S. Aubry and G. Andr\'{e}, {\it Analyticity breaking and Anderson localization in incommensurate lattices}, Ann. Isr. Phys. Soc. {\bf 3}, 133 (1980).

\bibitem{gaah1} J. Biddle and S. Das Sarma, {\it Predicted Mobility Edges in One-Dimensional Incommensurate Optical Lattices: An Exactly Solvable Model of Anderson Localization}, Phys. Rev. Lett. {\bf 104}, 070601 (2010).

\bibitem{gaah2} C. Danieli, J. D. Bodyfelt, and S. Flach, {\it Flatband engineering of mobility edges}, Phys. Rev. B {\b 91}, 235134 (2015).

\bibitem{gaah3} T. Devakul and D. A. Huse, {\it Anderson localization transitions with and without random potentials}, Phys. Rev. B {\bf 96}, 214201 (2017).

\bibitem{gaah4} S. Gopalakrishnan, {\it Self-dual quasiperiodic systems with power-law hopping}, Phys. Rev. B {\bf 96}, 054202 (2017).

\bibitem{gaah5} F. A. An, R. J. Meier, and B. Gadway, {\it Engineering a Flux-Dependent Mobility Edge in Disordered Zigzag Chains}, Phys. Rev. X {\bf 8}, 031045 (2018).

\bibitem{gaah6} A. Purkayastha, A. Dhar, and M. Kulkarni, {\it Nonequilibrium phase diagram of a one-dimensional quasiperiodic system with a single-particle mobility edge}, Phys. Rev. B {\bf 96}, 180204(R) (2017).

\bibitem{mott} N. Mott, {\it The mobility edge since 1967}, J. Phys. C {\bf 20}, 3075 (1987).

\bibitem{exaah1} Y. E. Kraus, Y. Lahini, Z. Ringel, M. Verbin, and O. Zilberberg, {\it Topological States and Adiabatic Pumping in Quasicrystals}, Phys. Rev. Lett. {\bf 109}, 106402 (2012).

\bibitem{exaah2} M. Lohse, C. Schweizer, H. M. Price, O. Zilberberg, and I. Bloch, {\it Exploring 4D quantum Hall physics with a 2D topological charge pump}, Nature (London) {\bf 553}, 55 (2018).

\bibitem{reentrant1} S. Roy, T. Mishra , B. Tanatar , and S. Basu, {\it Reentrant Localization Transition in a Quasiperiodic Chain}, Phys. Rev. Lett. {\bf 126}, 106803 (2021).

\bibitem{reentrant2} C. Wu, J. Fan, G. Chen, and S. Jia, {\it Non-Hermiticity-induced reentrant localization in a quasiperiodic lattice}, New J. Phys. {\bf 23}, 123048 (2021).

\bibitem{reentrant3} X.-P. Jiang, Y. Qiao, and J.-P. Cao, {\it Mobility edges and reentrant localization in one-dimensional dimerized non-Hermitian quasiperiodic lattice}, Chin. Phys. B {\bf 30}, 097202 (2021).

\bibitem{reentrant4} Z.-W. Zuo and D. Kang, {\it Reentrant localization transition in the
Su-Schrieffer-Heeger model with random-dimer disorder}, Phys. Rev. A {\bf 106}, 013305 (2022).

\bibitem{reentrant5} A. Padhan, M. K. Giri, S. Mondal, and T. Mishra, {\it Emergence of
multiple localization transitions in a one-dimensional quasiperiodic lattice}, Phys. Rev. B {\bf 105} L220201 (2022).

\bibitem{reentrant6} H. Wang, X. Zheng, J. Chen, L. Xiao, S. Jia, and L. Zhang, {\it Fate of the reentrant localization phenomenon in the one-dimensional dimerized quasiperiodic chain with long-range hopping}, Phys. Rev. B {\bf 107}, 075128 (2023).

\bibitem{reentrant7} S.-Z. Li and Z. Li, {\it The multiple re-entrant localization in a phase-shift quasiperiodic chain}, arXiv:2305.12321 (2023).

\bibitem{reentrant8} S. Aditya, K. Sengupta, and D. Sen, {\it Periodically driven model with quasiperiodic potential and staggered hopping amplitudes: Engineering of mobility gaps and multifractal states}, Phys. Rev. B {\bf 107}, 035402 (2023).

\bibitem{reentrant9} V. Goblot et al, {\it Emergence of criticality through a cascade of delocalization transitions in quasiperiodic chains}, Nat. Phys. {\bf 16}, 832 (2020).

\bibitem{reentrant10} A. \v{S}trkalj, E. V. H. Doggen, I. V. Gornyi, and O. Zilberberg, {\it Many-body localization in the interpolating Aubry-Andr\'{e}-Fibonacci model}, Phys. Rev. Research {\bf 3}, 033257 (2021).


\bibitem{helec1} A.-M. Guo and Q.-F. Sun, {\it Enhanced spin-polarized transport through DNA double helix by gate voltage}, Phys. Rev. B {\bf 86}, 035424 (2012).

\bibitem{helec2} S. Sarkar and S. K. Maiti, {\it Localization to delocalization transition in a double stranded helical geometry: Effects
of conformation, transverse electric field and dynamics}, Phys. Rev. B {\bf 100}, 205402 (2019).

\bibitem{qfsung} A.-M. Guo and Q.-F. Sun, {\it Spin-dependent electron transport in protein-like single-helical molecules}, Proc. Natl Acad. Sci. USA {\bf 111}, 11658 (2014).

\bibitem{helec3} A. V. Malyshev, {\it DNA Double Helices for Single Molecule Electronics}, Phys. Rev. Lett. {\bf 98}, 096801 (2007).

\bibitem{helec4} A.-M. Guo and Q.-F. Sun, {\it Topological states and quantized current in helical organic molecules}, Phys. Rev. B {\bf 95}, 155411 (2017).

\bibitem{helec5} T.-R. Pan, A.-M. Guo, and Q.-F. Sun, {\it Effect of gate voltage on spin transport along $\alpha$-helical protein}, Phys. Rev. B {\bf 92}, 115418 (2015). 

\bibitem{def-ipr1} X. Li, X. Li, and S. D. Sarma, {\it Mobility edges in one-dimensional bichromatic incommensurate potential}, Phys. Rev. B {\bf 96}, 085119 (2017).

\bibitem{def-ipr2} M. Rossignolo and L. Dell'Anna, {\it Localization transitions and mobility edges in coupled Aubry-Andr\'{e} chains}, Phys. Rev. B {\bf 99}, 054211 (2019).

\bibitem{dna} R. G. Endres, D. L. Cox, and R. R. P. Singh, {\it Colloquium: The quest for high-conductance DNA}, Rev. Mod. Phys. {\bf 76}, 195 (2004).

\bibitem{dim1} M. Ja\l{}ochowski, T. Kwapi\'{n}ski, P. \L{}ukasik., P. Nita, and M. Kopciuszy\'{n}ski, {\it Correlation between morphology, electron band structure, and resistivity of Pb atomic chains on the Si(5 5 3)-Au surface}, J. Phys.: Condens. Matter {\bf 28}, 284003 (2017).

\bibitem{dim2} M. Kopciuszy\'{n}ski, M. Krawiec, R. Zdyb, and M. Ja\l{}ochowski, {\it Purely one-dimensional bands with a giant spin-orbit splitting: Pb nanoribbons on Si(553) surface}, Sci. Rep. {\bf 7}, 46215 (2017).

\bibitem{dim3}  G. I. Japaridze and E. Pogosyan, {\it Magnetization plateau in the $S=\frac{1}{2}$ spin ladder with alternating rung exchange}, J. Phys.: Condens. Matter {\bf 18}, 9297 (2006).

\bibitem{dim4} Ding, H. et al. {\it Tuning interactions between spins in a superconductor}, Proc. Natl. Acad. Sci. {\bf 118}, e2023837118 (2021).

\bibitem{dim5} M. Krawiec, M. Kopciuszy\'{n}ski, and R. Zdyb, {\it Different spin textures in one-dimensional electronic bands on Si(5 5 3)-Au surface}, Appl. Surf. Sci. {\bf 373}, 26 (2016).

\bibitem{dim6} M. Sauter, R. Hoffmann, C. S\"{u}rgers, and H. v. L\"{o}hneysen, {\it Temperature-dependent scanning tunneling spectroscopy on the Si(557)-Au surface}, Phys. Rev. B {\bf 89}, 075406 (2014).


\end{thebibliography}
\end{document}